\documentclass[sigconf]{acmart}

\AtBeginDocument{%
  \providecommand\BibTeX{{%
    \normalfont B\kern-0.5em{\scshape i\kern-0.25em b}\kern-0.8em\TeX}}}

\usepackage{enumitem}
\usepackage{subcaption}
\usepackage{caption}
\usepackage{flushend}
\usepackage{xspace}
\newcommand{\alphaval}[2]{{\small $p\,#1\,#2$}}

\newcommand{\ptps}[1]{\textsc{ptps#1}}
\newcommand{\llms}[1]{\textsc{vlms#1}}
\newcommand{\ptp}[1]{\textsc{ptp#1}}
\newcommand{\llm}[1]{\textsc{vlm#1}}
\newcommand{\qo}[1]{\emph{QO#1}}
\newcommand{\qi}[1]{\emph{QI#1}}
\newcommand{\qroi}[1]{\emph{QR-O/I#1}}
\newcommand{\qro}[1]{\emph{QR-O#1}}
\newcommand{\qri}[1]{\emph{QR-I#1}}

\newcommand{\ivtask}[1]{\textsc{task#1}}
\newcommand{\ivmethod}[1]{\textsc{method type#1}}
\newcommand{\cours}[1]{\emph{SituationAdapt#1}}
\newcommand{\cauit}[1]{\emph{UserCentric#1}}
\newcommand{\cadapt}[1]{\emph{SurfaceAdapt#1}}

\newcommand{\anova}[6]{{\small [$F_{#1,#2}$\,$=$\,$#3$, $p$\,$#4$\,$#5$]}}
\newcommand{\pval}[2]{{\small ($p\,#1\,#2$)}} 
\newcommand{\pvall}[2]{{\small $p\,#1\,#2$}} 

\newcommand \change[1]{{\textcolor{black}{#1}}}
\newcommand\delete[1]{}

\copyrightyear{2024}
\acmYear{2024}
\setcopyright{acmlicensed}\acmConference[UIST '24]{The 37th Annual ACM Symposium on User Interface Software and Technology}{October 13--16, 2024}{Pittsburgh, PA, USA}
\acmBooktitle{The 37th Annual ACM Symposium on User Interface Software and Technology (UIST '24), October 13--16, 2024, Pittsburgh, PA, USA}
\acmDOI{10.1145/3654777.3676470}
\acmISBN{979-8-4007-0628-8/24/10}

\newcommand{\projname}{SituationAdapt\xspace}
\newcommand{\projectname}{\projname}
\newcommand{\framework}{system}

\begin{document}

\title{\projectname: Contextual UI Optimization in Mixed Reality with Situation Awareness via LLM Reasoning}

\author{Zhipeng Li, Christoph Gebhardt, Yves Inglin, Nicolas Steck, Paul Streli, and Christian Holz}
\affiliation{%
  \institution{Department of Computer Science, ETH Z\"urich}
  \city{Zurich}
  \country{Switzerland}
}


  
  
  

\renewcommand{\shortauthors}{Li et al.}



\begin{abstract}

Mixed Reality is increasingly used in mobile settings beyond controlled home and office spaces.
This mobility introduces the need for user interface layouts that adapt to varying contexts.
However, existing adaptive systems are designed only for \emph{static} environments.
In this paper, we introduce \emph{\projectname}, a system that adjusts Mixed Reality UIs to real-world surroundings by considering environmental and social cues in shared settings.
Our system consists of perception, reasoning, and optimization modules for UI adaptation.
Our perception module identifies objects and individuals around the user, while our reasoning module leverages a Vision-and-Language Model to assess the placement of interactive UI elements.
This ensures that adapted layouts do not obstruct relevant environmental cues or interfere with social norms. 
Our optimization module then generates Mixed Reality interfaces that account for these considerations as well as temporal constraints.
For evaluation, we first validate our reasoning module's capability of assessing UI contexts in comparison to human expert users.
In an online user study, we then establish \projname's capability of producing context-aware layouts for Mixed Reality, where it outperformed previous adaptive layout methods.
We conclude with a series of applications and scenarios to demonstrate \projname's versatility.

\end{abstract}



\begin{teaserfigure}
  \vspace{-2mm}%
  \includegraphics[width=\textwidth]{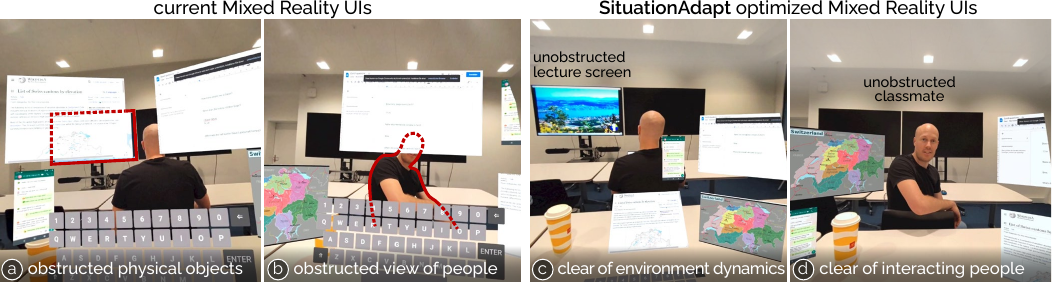}\vspace{-2mm}
  \caption{\projectname is an optimization-based adaptive UI system that reconciles Mixed Reality layouts with shared real-world spaces.
  Previous layout adaptations do not consider the situational context, such as (a)~if a shared display is on/off or (b)~if a classmate is facing the user.
  Our computational pipeline identifies these and other characteristics and adapts Mixed Reality layouts with situational awareness, such that here (c)~UIs stay clear of the video playback and~(d) the talking classmate.\vspace{1mm}}
  \Description{A four-panel image showcasing different computer setups in an educational setting with various applications open for multitasking during lectures or work sessions. Panel (a): Dual Monitor Setup. The left screen displays a lecture video. The right screen shows a lecture screen with various graphs and text. An overlay of a keyboard and two hands typing is visible at the bottom. Panel (b): Single Monitor with Multiple Windows. One large monitor with several open windows. A red-highlighted window indicates it’s being focused on by the user. The window labeled “classmate” is visible. Overlay of hands typing on a keyboard. Panel (c): Central Monitor with Data Analysis. A large central monitor displays a colorful map and data analysis. Smaller windows surround the central display, containing additional text and graphics. The term “lecture” appears at the top right corner. A person’s head is visible from behind, looking at the screen. Panel (d): Another Single-Monitor Setup. One window is in focus, with several others in the background. The focused window has ``SituationAdapt'' written at the top left corner. Another window labeled “classmate” is also visible. Personal items like notepads, pens, and a container (possibly holding food or drink) are in front of the monitor.}
  \label{fig:teaser}
\end{teaserfigure}


\maketitle

\section{Introduction}
Mixed Reality (MR) devices are becoming increasingly more mobile, which indicates a future where they will be commonplace and can be used in shared public and private spaces.
These can range from shared airplane~\cite{kari2023handycast} or train compartments, offices, and coffee shops to living rooms, kitchen areas, entire buildings~\cite{cheng2019vroamer} or public spaces~\cite{yang2019dreamwalker}---similar to the environments where we commonly use smartphones, tablets, and laptops today.

Unlike user interfaces (UIs) on traditional screen devices, however, MR UIs transcend device boundaries;
they can seamlessly blend into the user's physical surroundings and overlay parts of the real world.
Adapting and reconciling virtual layouts with physical surroundings for MR use is a challenging task.
Previous research has optimized MR UIs for proximity with semantically similar physical objects~\cite{semanticadapt2022cheng} or leveraged physical affordances of the user's surroundings to facilitate efficient interaction~\cite{cheng2023interactionadapt}.
These adaptations have so far focused on the real-world objects and surfaces within the user's reach inside their (personal) workspace, often assuming static environments during use.

In \emph{shared} spaces, social norms become meaningful during interaction.
Therefore, MR layouts must additionally conform to the social situations and dynamic environmental conditions that can take place in such environments.
Previous studies highlight this need in their investigation of MR use in shared spaces (e.g.,~\cite{medeiros2022shielding,grubert2016towards,williamson2019planevr,whereputit2022luo}).
The results of these studies indicate that users find it crucial for MR UI layouts to consider factors such as the functionality of objects in their surroundings, the social appropriateness of element placement, the effects of UI positioning on health \& safety, and maintaining the visual appeal of the physical environment.

In this paper, we propose \emph{\projname}, a system that optimizes MR layouts for situational social and environmental factors.
Our \framework{} consists of perception, reasoning, and optimization modules to reconcile adapted MR UIs with real-world environments and conform to social norms and dynamic conditions.

\subsection*{Adapting UIs to Shared Real-World Settings}

\autoref{fig:teaser} illustrates the challenge of situation-aware UI adaptation at using a lecture scenario.
While a UI element can be suitably positioned in front of a classmate as he faces away from the MR user (\autoref{fig:teaser}c), placing the same widget in front of his face as he faces or even interacts with the MR user is intrusive, as it not just impedes personal communication but also renders direct interaction with the MR UI inappropriate (\autoref{fig:teaser}b).
Likewise, UI elements may be placed in front of a physical screen, since they do not obstruct any information (\autoref{fig:teaser}d).
When the screen comes on, however, the virtual element occludes potentially meaningful content (\autoref{fig:teaser}a).

\projname reconciles the layout of virtual UI elements with real-world conditions to ensure appropriate placement using the three modules of our system.
This avoids intrusiveness and maintains considerate functionality in dynamic environments.

Our \emph{perception module} identifies objects and people in the physical environment through a real-time object detection network while simultaneously reconstructing a 3D map of the user's surroundings.
The module then segments identified objects and people from the 3D map to extract them as input into our optimization scheme.

Our \emph{reasoning module} leverages a Vision-and-Language Model (VLM) to evaluate the potential placement of UI elements within a shared social space.
Based on prior research, we designed a prompt to consider factors such as functionality, aesthetics, social acceptability, and health \& safety.
Because observing a UI element that occludes part of a shared space has different implications for these factors than a user's direct interactions with that UI element, we separately query the VLM for \emph{overlay suitability} and \emph{interaction suitability}.
From the VLM response, we extract ratings to inform a goodness function for UI element placement that considers relevant environmental cues as well as social norms. 

Finally, our \emph{optimization module} processes the 3D bounding boxes of objects and people in the physical environment and the associated suitability ratings for overlaying content for display or interaction.
From these inputs, the module generates layouts of MR UIs that account for environmental and social aspects of shared spaces.
We propose two novel optimization terms for interactive MR adaptation \delete{(using AUIT~\cite{evangelista2022auit})} that model the suitability for overlaying and interaction.
Integrated into our real-time system, these terms optimize MR UIs for suitable viewing and interaction given the current shared physical environment.

We evaluate the efficacy of \projectname in two studies: an online survey to evaluate our reasoning module and an in-situ user evaluation to evaluate our end-to-end system.
In the online survey, we validate if the underlying VLM of our reasoning module judges the context of shared spaces similar to pre-screened, experienced MR users.
We collected ratings from 42 participants and 42 VLM instances, evaluating 64 areas of interest within 18 diverse scenarios.
The results of the survey indicate that, across scenarios, VLMs achieved comparable ratings to participants for both, overlay and interaction suitability.

We then conducted a user study to compare \projectname's optimized layouts with those of two representative baseline methods that do not account for shared spaces.
Participants perceived \projectname{}'s MR layouts to more suitably overlay UI elements onto the physical environment and position them more appropriately for interaction within the context of a shared social space.
Participants also expressed a strong preference for the layouts generated by \projectname{} compared to those from baseline methods.
Finally, we demonstrate \projname's applicability across two scenarios within diverse shared spaces.

\subsection*{Contributions}

We make the following contributions in this paper.

\begin{itemize}[leftmargin=*]
    \item an optimization-based end-to-end \framework{} that considers aspects of MR use in shared spaces in the optimization of MR layouts through an VLM-based reasoning component.
    Our approach can adapt UI element placements while taking into account their impact on, for instance, occluding real-world objects' functions, social appropriateness, health \& safety, and the aesthetic appeal of the surroundings.
    
    \item a crowd-sourced survey study ($N=42$) that demonstrated that our VLM-based reasoning module judges the context of shared spaces not different than experienced MR users.
    
    \item an empirical study that compared \projectname\ to two baseline approaches ($N=12$), showing that our approach generated layouts that participants preferred and rated more appropriate for shared spaces than the baseline layouts.
    
    \item two proof-of-concept scenarios that integrate our \framework{} to adapt MR layouts to the situational context of a shared space.
\end{itemize}

\section{Related work}

\projname is related to Mixed Reality usage in shared settings, adaptive layout systems for Mixed Reality, and the use of large language models in HCI.

\subsection{Mixed Reality in shared spaces}
Researchers have been exploring the effect of environmental and social dynamics of shared spaces on the use of MR devices \cite{grubert2016towards} and interaction in MR.
Transportation settings have been studied in depth~\cite{mcgill2020challenges, li2021rear, mcgill2019virtual}, where shared surroundings demand socially acceptable and safe interaction~\cite{williamson2019planevr}, especially given the lack of space for expansive input~\cite{kari2023handycast}.
\citeauthor{medeiros2022shielding} studied the layout of MR interfaces in shared transit contexts, including vehicles and trains, and \delete{concluded several considerations when}\change{identified important aspects for} using VR in shared spaces: social etiquette, spatial affordance, and safety~\cite{medeiros2022shielding}.

Other works have considered multiple users and bystanders within shared environments, such as for collaboration scenarios with multiple MR users~\cite{whereputit2022luo} or individual MR users and projected augmented reality~\cite{gugenheimer2017sharevr}.
\citeauthor{o2023re} explored the MR interactions with bystanders, reporting the need for socially intelligent bystander awareness systems~\cite{o2023re}.

While there are multiple factors influencing the experience of MR users and bystanders in shared spaces, it is hard to comprehensively model them in a computational manner.
Our work leverages the reasoning capabilities of modern VLMs to understand the context of shared spaces and integrate inferred contextual information into an optimization scheme.

\vspace{-3mm}
\subsection{Adaptive Mixed Reality Interfaces}
Prior research has explored adapting MR interfaces to various contextual factors including the user or their state, the task, as well as the physical environment.

One essential focus of MR adaptive user interfaces is environment-driven adaptation~\cite{hettiarachchi2016annexing, realitycheck2019hartmann, niyazov2023user, evangelista2022auit}.
Employing geometry-based approaches, researchers have suggested aligning virtual contents with the physical surroundings (e.g., Flare~\cite{flare2014gal}, Optispace~\cite{fender2018optispace}, TapID~\cite{meier2021tapid}, TapLight~\cite{streli2023structuredlightspeckle}).
\citeauthor{lages2019walking} dynamically adapted virtual elements to physical windows and walls when the user was walking~\cite{lages2019walking}.
\citeauthor{scalar2022qian}'s decision tree-based strategy adapted AR interfaces to new environments while keeping the semantic relationships between virtual and physical elements from the previous layout~\cite{scalar2022qian}.
\citeauthor{kari2021transformr}'s TransforMR method detected people and dynamic elements in MR scenes and substituting them with alternative avatars or objects through diminishing, thereby imbuing the physically plausible behavior of the original objects onto the synthetic replacements~\cite{kari2021transformr}.
Asynchronous Reality dynamically diminished real-world objects to preserve the impression of the user's surroundings at one point in time when their state changed~\cite{fender2022asyncreality}.
SemanticAdapt included the semantic relationship between virtual and physical objects with other factors such as temporal consistency, occlusion, and proposed an integer-programming-based optimization approach to obtain the adaptive interface~\cite{semanticadapt2022cheng}. 
Our previous UI adaptation method InteractionAdapt~\cite{cheng2023interactionadapt} additionally focused layout optimization on situated affordances such as physical surfaces and obstacles with empirically quantified benefits for interaction~\cite{cheng2022comfortable,luong2023controllersbarehands} to provide passive haptic feedback and rest for optimized MR interaction during prolonged tasks between within-reach and far-away objects while accounting for physical obstacles that prevent input.

Other approaches investigated the adaptation of MR interfaces to the user's state~\cite{visualattentionoptimization2019alghofaili,lu2022exploring,veras2021elbow}.
\citeauthor{gazeRL2019Gebhardt} learned to display labels of virtual elements based on users' gaze interactions with the VR environment \cite{gazeRL2019Gebhardt}.
\citeauthor{contextAwareMR2019lindlbauer} optimized virtual elements' visibility, level of detail and placement based on the estimation of users' cognitive load from pupil dilation~\cite{contextAwareMR2019lindlbauer}.
\citeauthor{xrgonomics2021belo}~\cite{xrgonomics2021belo} and \citeauthor{ergo2017maurillo}~\cite{ergo2017maurillo} further optimized virtual interfaces for ergonomics with rule-based estimation~\cite{mcatamney1993rula}.

Newer work proposed a Pareto-optimal method to achieve a balance between competing objectives for MR UI adaptation \cite{paretooptimal2023johns} or introduced a tool to help researchers design new MR interfaces in various contexts based on previously collected MR UIs \cite{cho2024minexr}.

While research on adaptive MR interfaces explored numerous factors and settings, we are the first to adapt MR layouts to shared spaces considering factors such as 'social acceptability' and 'health \& safety'. Our end-to-end system recognizes relevant cues in shared social settings and can optimize a MR UI accordingly.

\subsection{LLMs in HCI}
Recent advancements in Large Language Models (LLMs) have created widespread excitement across research disciplines, exploring their potential application to various tasks.
In HCI, research has explored LLMs for tasks such as writing \cite{chung2022talebrush, gero2022sparks}, learning \cite{baidoo2023education, kung2023performance}, and programming \cite{vaithilingam2022expectation, chen2021evaluating, pearce2023examining}.
Other works explored using LLMs to facilitate information retrieval \cite{jiang2305graphologue}, manage information with multilevel abstraction \cite{suh2023sensecape} and synthesize scholarly literature \cite{kang2023synergi}.

Most similar to our work, is research that uses LLMs to simulate participants of a user study.
\citeauthor{hamalainen2023evaluating} utilized GPT3 to generate open-ended responses about video game experiences and found that the LLM produced answers comparable to those of human participants \cite{hamalainen2023evaluating}.
\citeauthor{schmidt2024simulating} found out that one might obtain artificial answers when using LLMs to simulate survey participants, but also highlight that LLMs give unanticipated responses that offer new insights and help to discover pitfalls in the survey design \cite{schmidt2024simulating}. To validate LLM responses, they suggest to combine small-scale user studies with large-scale user simulation.

Following their suggestion, we validate the feasibility of our approach of utilizing a VLM to rate the suitability of placing virtual elements in shared social spaces with an online survey.
In this evaluation, we compared VLM responses to those of experienced MR users in terms of understanding the context of shared spaces.


\section{Adaptive MR for Shared Spaces}
\label{sec:context_shared_spaces}
We define the factors to consider when developing adaptive MR layout approaches for shared spaces. 
By reviewing the aspects that previous studies consistently highlighted as crucial, we derive the following four key factors.

\begin{figure*}[t]
  \centering
  \includegraphics[width=\textwidth]{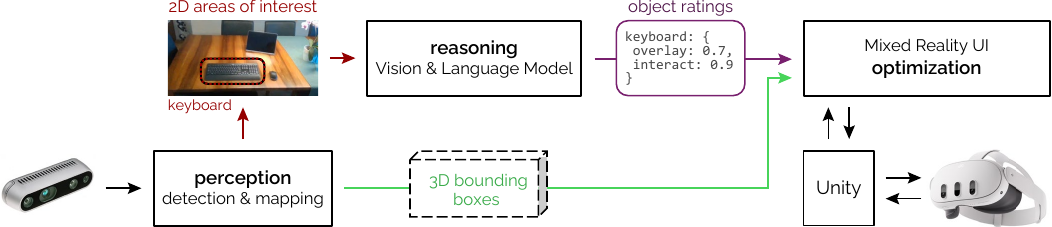}
  \caption{Schematic overview of \projectname's system.
  Our perception module recognizes 2D areas of interest in the environment and computes 3D bounding boxes of the respective objects. Our reasoning module takes the areas as input and leverages a VLM to rate their overlay- and interaction suitability. Unity then assigns these ratings to the respective 3D bounding boxes and our optimization module adapts MR UIs accordingly.}
  \Description{A flowchart depicting various elements of a user interface optimization process. On the left, there’s an image labeled ‘2D areas of interest’ showing a keyboard with a red overlay. Connected to it is a box labeled ‘reasoning VLM.’ Below this is another image of a device with the label ‘perception detection \& mapping.’ These two boxes are connected to a central box with dashed lines labeled ‘3D bounding boxes.’ To the right, there are two more boxes; one labeled ‘object ratings’ with subtext ‘keyboard: {overlay: 8.7, interact: 9.9}’ and another box titled ‘MR UI optimization’ pointing towards an image of a VR headset next to the word ‘Unity.’ The flowchart illustrates the process from real-world input through various computational assessments leading to mixed reality user interface optimization using Unity software.}
  \label{fig:overview}
\end{figure*}

\begin{enumerate}[leftmargin=0.75cm]
  \item[\textbf{F}] Functionality: UI elements hinder the functionality of a physical object (e.g., cup, laptop, display)~\cite{whereputit2022luo, medeiros2022shielding}.
  
  \item[\textbf{A}] Aesthetics: UI elements impair the visual appeal of the physical surroundings~\cite{medeiros2022shielding}.
  
  \item[\textbf{S}] Social acceptability: looking at or directly interacting with UI elements is \change{considered} socially inappropriate \change{by bystanders}~\cite{williamson2019planevr, medeiros2022shielding, grubert2016towards, medeiros2023surveying}.
  
  \item[\textbf{H}] Health \& Safety: UI elements occlude safety critical information or lead to sanitation issues during interaction~\cite{williamson2019planevr, wilson2023lack}.
\end{enumerate}

\delete{
Furthermore, 
we respect that these factors depend on whether a user is solely looking at a UI element or directly interacting with it.
For example, while it may be socially acceptable for a user to glance at the map widget in Figure 1, directly interacting with it could be deemed inappropriate, as it might distract other students attending the lecture.
}

\delete{
Therefore, we model this suitability type into our MR layout adaptation strategy for shared spaces and differentiate between \emph{overlay suitability} and \emph{interaction suitability}. 
}
\delete{
\emph{Overlay suitability} evaluates whether integrating virtual elements into the user's per\-ception of the real world violates or supports the previously defined factors (\textbf{FASH}).
}
\delete{
\emph{Interaction suitability} evaluates the appropriaten\-ess of interact with a UI element subject to \textbf{FASH}.
}

\change{Furthermore, we respect that whether a user is soly observing a UI element or directly interacting with it can impact the FASH factors differently. 
For instance, while it may be socially acceptable for a user to glance at the map widget in \autoref{fig:teaser}, direct interaction with it could be inappropriate, as it might distract other students attending the lecture. 
Similarly, placing a widget above the back of a passenger’s head on a bus is suitable for observation but may be socially inappropriate for interaction, as it could lead to physical contact with the person's head. 
Therefore, we model suitability using two distinct scores: one for when a UI widget is being observed (overlay suitability) and another for when it is being interacted with (interaction suitability).
}

\change{
To address these differences, we define \emph{Overlay suitability} as the cumulative appropriateness of the \textbf{FASH} factors when a UI widget is being looked at and \emph{Interaction suitability} as the cumulative appropriateness of the \textbf{FASH} factors when a UI widget is interacted with.
We use these scores define the output of the VLM.
With this formulation, the VLM can balance the impact of potentially conflicting \textbf{FASH} factors on placement suitability.
This approach is more robust than treating each FASH factor as a separate objective term in an optimization scheme and relying on weight tuning to balance conflicting factors.
}

\section{Method}


\projectname adjusts MR UIs to real-world conditions by considering social cues in shared settings.
\autoref{fig:overview} provides an overview of our \framework: a perception module recognizes objects and people around the user and fits 3D bounding boxes around them. 
Our reasoning module leverages a VLM to evaluate the suitability of scene locations to accommodate UI elements for display and/or interaction, ensuring that widgets do not obstruct relevant real-world cues or interfere with social norms.
Our optimization scheme then uses these ratings as well as the 3D bounding boxes of the respective objects and people as input and generates MR interfaces that account for these aspects.

Below, we explain the operation of our modules.
First, we discuss the functionalities and mechanisms of the perception module and the reasoning module.
Finally, we detail the formulation of our optimization scheme.

\subsection{Perception of surroundings}
The perception module receives RGBD frames as input and provides semantically annotated 2D- and 3D bounding boxes of areas of interest as output.
Areas of interest characterize the objects and people that were found in the real-world surroundings of the MR user, defined by the typical categories recognized by real-time object detection networks.

Modern MR headsets, such as Meta Quest 3, posses sophisticated inside-out tracking capabilities that can track the physical environment and even the dynamic user body. 
Recent developments indicate that these headsets will soon also have the capability to understand the 3D space around the user~\cite{meta2024scenescript}.
While these advancements already exist or are within reach, SDKs of current MR headsets do not make them available for developers.
For this purpose, we developed a custom perception module (Section~\ref{sec:perception_module}).

\vspace{-3mm}
\subsection{Reasoning about placement suitability}
\label{sec:suitabiltiy-identification}
The reasoning module takes RGB images annotated with the areas of interest as input (\autoref{fig:scenario_examples} illustrates examples of such images).
For these images, we then query a VLM to rate the overlay- and interaction suitability for hypothetical UI elements positioned within box on a scale from from 1 ('unsuitable') to 5 ('suitable').
We start this evaluation by setting the context of the VLM, explaining what we mean with overlay- and interaction suitability of Mixed Reality UIs.
We further prime it with the factors we derived to be important in the context of using Mixed Reality in shared spaces (Section~\ref{sec:context_shared_spaces}).
The comprehensive context prompt is detailed in \autoref{app:context-prompt}.

\delete{
As initial tests have revealed that the VLM's ratings diverge from user ratings, we additionally fine-tune it.
Therefore, we provide the VLM with previously rated images and their respective ratings.
}
\change{As initial tests revealed discrepancies between the VLM's ratings and user ratings, we have incorporated previously user-rated images and their respective ratings into the context of the VLM.}
\delete{More precisely, for each designated area within an image, we prompt it with the median rating and the standard deviation of a population of users}\change{More precisely, for each designated area within one of the user-rated images, we prompt the VLM with the median and the standard deviation of the ratings of a group of users.
Using this context, we then query the VLM to rate overlay- and interaction suitability of a previously unseen image.}
Our tests have shown that this process increases the model's understanding of how users would rate situations and helps the VLM to align its ratings with those of users.

Finally, we query overlay- and interaction suitability for the areas of an unseen image with the following prompt: "Please rate the suitability of overlaying/directly interacting with a virtual UI element on each area in this image. The acquired ratings are forwarded to Unity and the optimization module.

\subsection{Optimizing the MR UI layout}
\label{sec:optimization}

We base our optimization module on the AUIT toolkit \cite{evangelista2022auit}.
The general form of the objective function of AUIT is defined as 
\begin{equation}
    Q = \sum_{i=1}^{V}\sum_{j=1}^{O} w_{ij}c_{ij}(\textbf{x})
\end{equation}
where $V$ is the set of virtual elements and $O$ is the set of objectives, both accompanied by corresponding weights $w$ and cost functions $c$.
$\textbf{x}$ is the decision vector comprising configuration parameters for all UI elements, optimized to minimize $Q$.
In our optimization scheme, we utilize five pre-defined objective terms of AUIT: Occlusion, Look towards, Distance, Field of view, Constant view size (for details see the original paper).

To generate MR layouts that are sensitive to situations in shared spaces, we propose two new terms to model \emph{overlaying suitability} and \emph{interaction suitability}.
Both terms take the detected 3D bounding boxes and the normalized 5-point suitability ratings (scaled between 0 and 1) as input.
\change{
In contrast to the occlusion term in AUIT, which models the appropriateness of UI widgets being occluded by other UI widgets or physical objects, our terms consider the occlusion of real-world objects and people by virtual content during display and interaction, enabling situation-aware MR UIs.
}

To compute the \emph{overlaying suitability} cost function, we rasterize each virtual element at an equal interval and cast a set of rays $R$ from the users' point of view to each point within the grid.
For each ray $r$, we then obtain a set of hit points $H(r)$ that constitutes the positions where the ray hit a 3D bounding box.
Based on these sets, we can now compute the cost function $c_{v, over}$ for overlaying a virtual element $v$ as,

\begin{align}
    c_{v, over} &= \sum_{r}^{R} \sum_{h}^{H(r)} p_{b} e^{-5 d_{h}}~, \\
    d_{h} &= \frac{||h - c_b||}{0.5 d_b} \nonumber
    \label{eq:overlaying}
\end{align}

where $h$ is a hit point, $c_b$ the center of the bounding box it hit, $d_b$ the length of the box's diagonal, and $d_{h}$ the respective normalized distance.
We employ an exponential function to implement a higher penalty when the hit point is close to the center of the bounding box.
The term $p_{b}$ is the penalty of overlaying a bounding box $b$ and is calculated as,

\begin{equation}
    p_{b} = \begin{cases}
        0.5 - o_{b},& o_{b} \leq 0.5 \\
    0,              & \text{otherwise}
    \end{cases}
\end{equation}

where $o_{b}$ is the suitability score for overlaying the bounding box $b$.
The term penalizes unsuitable boxes ($o_{b} \leq 0.5$), considering all others as suitable by default.

Similarly, we adopt the same grid-based ray casting procedure to compute the cost function for \emph{interaction suitability} as

\begin{equation}
    c_{v, inter} = \sum_{r}^{R} \sum_{h}^{H(r)} f_{v}  (0.5 - i_b) e^{-5 d_{h}}
    \label{eq:interaction}
\end{equation}

where $i_b$ is the interaction suitability score of bounding box $b$.
$f_v$ represents how frequently a virtual element $v$ is interacted with and hence needs to be penalized more in the context of this cost term (similar to respective terms in \cite{cheng2023interactionadapt, semanticadapt2022cheng, contextAwareMR2019lindlbauer}).   
Intuitively, this term encourages placing virtual element over physical bounding boxes which are suitable for interaction by introducing a negative penalty when $i_b$ is larger than 0.5.

\section{Implementation}
We now outline the implementation of each of \projname's modules.
Websockets facilitate the communication between them.
Our entire pipeline runs on an Intel Core i7-12700K with a NVIDIA GeForce GTX 1050 Ti and 32 GB of RAM.

\subsection{Perception module}
\label{sec:perception_module}
\begin{figure}[t]
  \centering
  \includegraphics[width=\columnwidth]{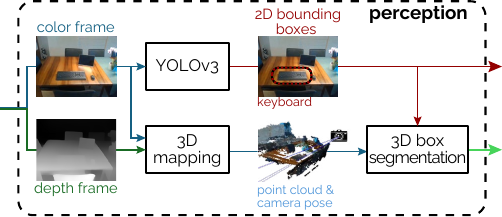}
  \caption{Our implementation of the perception module. Based on color- and depth frames of an RGBD camera, a 3D mapping stage reconstructs the camera position and the surroundings of the user as point cloud. An object detection node computes semantically annotated 2D bounding boxes. The last stage segments 3D bounding boxes based on the 2D ares, the point cloud and the camera position.}
  \Description{A diagram illustrating the process of object perception in computer vision. On the left, there’s an image labeled ‘2D areas of interest’ showing a keyboard with a red overlay. Connected to it is a box labeled ‘reasoning VLM.’ Below this is another image of a device with the label ‘perception detection & mapping.’ These two boxes are connected to a central diagram showing a ‘point cloud & camera pos,’ which leads to ‘3D box seg.’ for object segmentation, as part of the perception phase in computer vision.}
  \label{fig:perception_module}
\end{figure}

%
The perception module aims to identify areas of interest as 2D- and 3D bounding boxes, serving as input for the reasoning- and optimization module.
Our system utilizes the headset's inside-out tracking to maintain accurate positioning within the MR environment.
We transform bounding boxes from our perception module to Unity using a manually specified transformation matrix. 
As our system adapts MR UI layouts at a situational change of a shared space, the perception module is manually triggered when such a change happens.
We implemented the module within the Robot Operating System (ROS) where we ran separate ROS nodes for its three stages: 3D mapping, object detection, and 3D bounding box segmentation (see \autoref{fig:perception_module} for an overview). 
Depth and color frames of an Intel RealSense D435 RGB-depth camera serve as input to the module. 
In the following, each stage is briefly explained.

\subsubsection{3D mapping}
We utilize the RTAB-Map implementation of the Simultaneous Localization And Mapping (SLAM) algorithm \cite{noauthor_rtab-map_2023}
to fuse RGB- and depth frames into a global 3D map of the surroundings.
The resulting point cloud and camera position are forwarded to the 3D bounding box segmentation stage.

\subsubsection{Object detection}
We use YOLOv3~\cite{redmon_yolov3_2018} to detect objects and people in the scene. 
It takes the color image as input and outputs a category, confidence, and bounding box for each detected object or person. The annotated 2D bounding boxes are forwarded to the reasoning module as well as the 3D bounding box segmentation.

\subsubsection{3D bounding box segmentation}
\label{sec:sense-making}
%
In this stage, we reproject the corners of the annotated 2D bounding boxes into the mapped 3D scene to create a frustum. This frustum is then transformed from camera to world coordinates, and its signed distance function is computed for point selection within it. Points not visible due to occlusion are removed using the hidden point removal algorithm \cite{katz_direct_2007}. Finally, the DBSCAN algorithm \cite{ester_density-based_1996} clusters the frustum's point cloud, a bounding box is built around the largest cluster, and points from other clusters are eliminated.
Should a previously identified bounding box closely match the new one, it is replaced by the updated version. 
Conversely, if no similar bounding boxes are found, the new detection is incorporated into the scene as a separate entity. 
For each frame, all recognized bounding boxes are transmitted to Unity where a transformation is performed to convert them into Unity's coordinate system.

\subsection{Reasoning module}
We utilize the GPT4 Vision 2024-02-15-Preview model of Azure OpenAI as our VLM and access it via its Python API.
As Azure AI Services lack the capability to fine-tune models through direct training on image data, we employ few-shot learning to \delete{fine-tune}\change{provide} our VLM \change{with information about previously rated scenarios}. 
This involves integrating example images and corresponding ratings, as described in Section~\ref{sec:suitabiltiy-identification}, into its context prompt (see \autoref{app:context-prompt} for the specific prompt).
We prompt the VLM to provide its answer in the format: Area <area index>: <score>, <reason>.
The acquired ratings are transferred into Unity and assigned as properties to the respective 3D bounding boxes.

\subsection{Unity \& optimization module}
We implement our system for the Meta Quest~3 using Unity~2021.
To implement our MR UI optimization module, we leveraged AUIT~\cite{evangelista2022auit}, a toolkit to create adaptive Mixed-Reality applications.
The toolkit interacts directly with Unity, utilizing Unity GameObjects and properties as input to its optimization.

\section{Reasoning Validation}
\label{sec:reasoningvalidation}
Our pipeline is built on the hypothesis that \projectname can adequately understand the situational context of a shared space.
To evaluate this assumption, we conducted an online survey to compare the judgment of different situations in shared social spaces of \projectname with those of experienced MR users.

\subsection{Survey design}

Our survey sought to learn how \projectname and experienced MR users judge the suitability of overlaying and directly interacting with virtual UIs in various scenarios and shared social spaces. 
In instances where parts of these scenes were deemed unsuitable for either, we further tried to discern which of factors we identified as critical (\textbf{FASH}) underlies the judgment.
Thus, prior to starting the survey, we explained the suitability terms and the factors to participants. 
In addition, we showed two videos displaying the first-person view of a MR user in a share space.
After the introduction, participants continued answering demographic questions before starting with the main part of the survey.

\subsubsection*{Scenarios.}

\begin{figure*}
    \centering
    \begin{subfigure}{.3\textwidth}
        \centering
        \includegraphics[width=.95\linewidth]{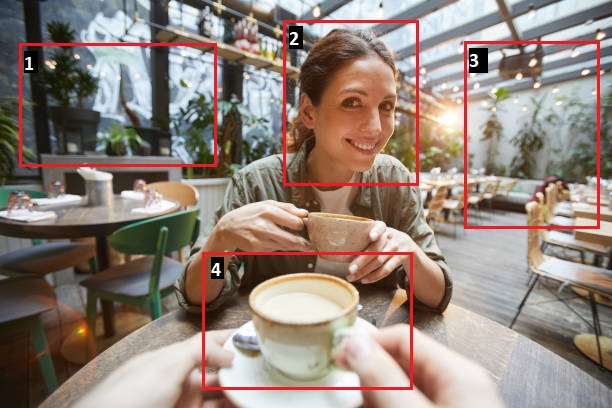}
    \end{subfigure}
    \begin{subfigure}{.3\textwidth}
    \centering
        \includegraphics[width=.95\linewidth]{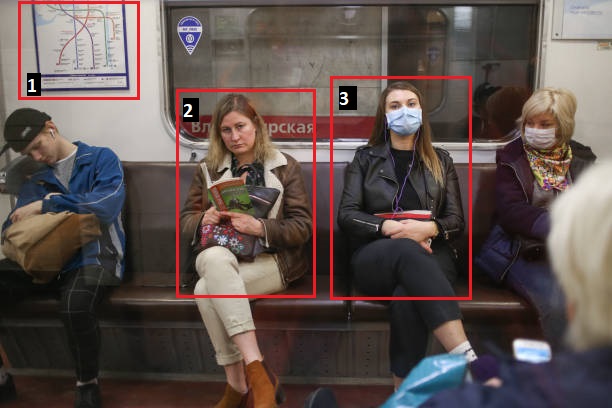}
    \end{subfigure}
    \begin{subfigure}{.3\textwidth}
    \centering
        \includegraphics[width=.95\linewidth]{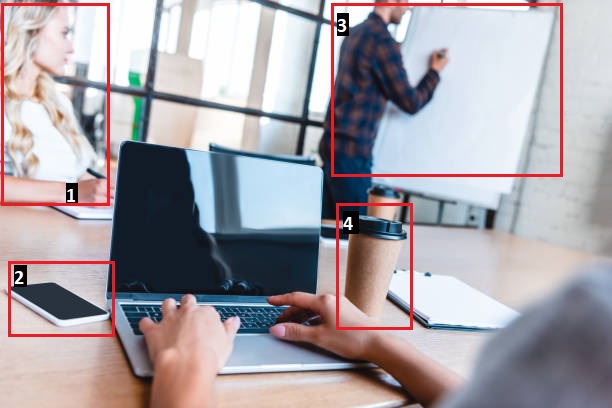}
    \end{subfigure}    
    \caption{Our survey covered these and other scenarios. Participants rated the overlay and interaction suitability for each area.}
    \Description{A triptych of images depicting various scenes with red rectangles highlighting specific areas. The first image shows a close-up of a person’s hand holding a cup of coffee with a red rectangle around the cup. The second image features three individuals seated at a table in what appears to be a café, with two red rectangles highlighting what seems to be digital devices on the table. The third image is an over-the-shoulder view of someone working on a laptop, with the screen and part of the keyboard enclosed in a red rectangle.}
    \label{fig:scenario_examples}
\end{figure*}

The main part of our survey consisted of 18 scenarios participants had to judge. Each scenario is a photo taken from first-person view of a hypothetical MR user. 
\delete{In each photo, we highlighted certain areas with bounding boxes.}
\change{In each photo, we manually designed bounding boxes to create challenging scenarios for the VLM to analyze, following this rule: Placing a widget within the bounding box must affect one or more FASH factors (e.g., occluding or being near a person the user is talking to, blocking an important safety-related sign).}
Participants had to rate these areas for their overlay- and interaction suitability.
We selected these scenarios to capture a wide variety of situations \delete{that happen in shared spaces.} \change{, including typical shared spaces (restaurant, airplane, home, office), social contexts (alone, with friends, strangers) and tasks (recreation, work).}
\autoref{fig:scenario_examples} shows three scenarios presented in our survey with the respective areas (illustrated through bounding boxes) participants had to judge.

\subsubsection*{Questions.}

For each of the highlighted areas of a scenario, participants had to answer four questions. 
First, they were asked to rate the suitability of overlaying a virtual UI element on each area (\qo : ``Please rate the suitability of overlaying a virtual UI element on each area in a Mixed Reality experience'').
Second, they should rate the suitability of directly interacting with envisioned virtual UI elements that were to be positioned in each area (\qi: ``Please rate the suitability of directly interacting with virtual UI elements displayed in each area. Note: All virtual elements are positioned within your arm's reach. If a virtual element covers a physical object, interacting with it means physically touching that object.'').
Responses to both questions were recorded using a 5-point Likert scale, with options ranging from ``Unsuitable'' to ``Suitable'' (1: ``Unsuitable'', 2: ``Somewhat unsuitable'', 3: ``Neutral'', 4: ``Somewhat suitable'', 5: ``Suitable'').
If participants selected `unsuitable' or `somewhat unsuitable' for an area in either question, they were asked to provide the reason (\qroi: ``If you selected `unsuitable' or `somewhat unsuitable', please select the primary reason for your choice.'').
The response options corresponded to the underlying factors outlined in the survey's introduction (Functionality, Social Acceptability, Health \& Safety, Aesthetics, Other with a text field to specify it).

\subsection{Participants}

We recruited 50 participants (16 female, 34 male), ages 22--50 (M=32, SD=9.1) from an online crowd-sourcing platform.
To guarantee a certain level of VR experience among participants, we screened them to ensure they used a VR device at least 6 times a month.
Of those, 13 participants reported using VR more than 15 times a month, 7 participants used it 11--15 times, and the remaining participants used it more than 6 times per month.
Participants also reported their frequency of using direct touch to interact in Mixed Reality: 3 participants reported daily use, 13 mentioned using it several times a week, 18 indicated they used it several times a month, and the remaining participants used it less frequently.
Participants completed the survey in 45\,min and received \pounds 6 as a gratuity.

We excluded participants that answered one third of our control questions wrong (more than 6 out of 20 control questions) as well as participants that gave extreme extreme median responses (1: ``Unsuitable'' or 5: ``Suitable'') with a standard deviation lower than one across all areas and images. 
Consequently, the data from 42 participants (\ptps{}) were used in the analysis.

\subsection{Generated suitability ratings}

To generate results with \projectname, we employed the identical scenarios and areas as for participants, and used the requests outlined in Section~\ref{sec:suitabiltiy-identification} to generate results with our perception module.
To ensure a matching sample size, we produced ratings from 42 distinct VLM instances, aligning the quantity of ratings between participants and \projectname. 
We split the scenarios in training- and test set and \delete{used}\change{add} the training set (9 out of 18) \delete{to fine-tune}\change{to the context of} VLM instances following the process described in Section~\ref{sec:suitabiltiy-identification}.
We then generated ratings for each question across the unseen scenarios, resulting in ratings from 42 VLM instances (\llms{}) across 9 scenarios with 3 or 4 areas each. 
For our analysis, this yielded a total of 1,344 ratings per condition (\llms{} and \ptps{}).

\subsection{Results}

The goal of our analysis was to determine if \projectname~ assesses overlay and interaction suitability of social scenarios similar to the population of experienced MR users.
Hence, we postulate the following null hypothesis:

\begin{enumerate}[leftmargin=*]
  \item[$H_0$] Instances of \llms{} provide overlay/interaction suitability ratings that deviate more extreme than those provided by individual \ptps{} in comparison to their broader population.
\end{enumerate}

\noindent
To analyze \qo\ and \qi, we employ bootstrap hypothesis testing~\cite{hall1991two, gentle2002}.
For each \ptp{} and \llm{}, we assess if their ratings significantly deviate from the rest of \ptps{} for every scenario and area (using the Mann-Whitney U test).
Across all 1764 bootstrap iterations, we count the percentage of instances where the ratings of an \llm{} diverge more often than those of a \ptp{} and normalize this count with the number of total comparisons, which determines the p-value \cite{hall1991two}.
For both questions, we can reject $H_0$ (\qo: \alphaval{<}{0.04}; \qi: \alphaval{=}{0.0}), indicating that \llms{} rate scenarios not significantly \emph{more} different to all \ptps{} than any individual \ptp{}.

\begin{figure}[b]
    \centering
    \begin{subfigure}{.3\columnwidth}
        \centering
        \includegraphics[width=.75\linewidth]{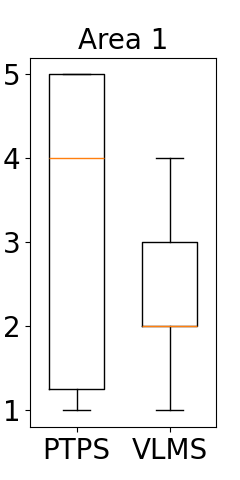}
        \caption{}
    \end{subfigure}
    \begin{subfigure}{.3\columnwidth}
    \centering
        \includegraphics[width=.75\linewidth]{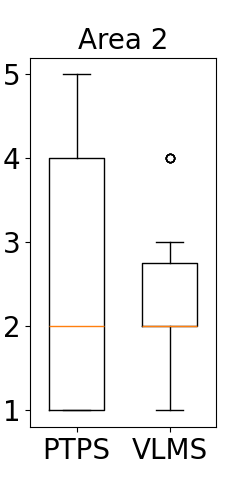}
        \caption{}
    \end{subfigure}
    \begin{subfigure}{.3\columnwidth}
    \centering
        \includegraphics[width=.75\linewidth]{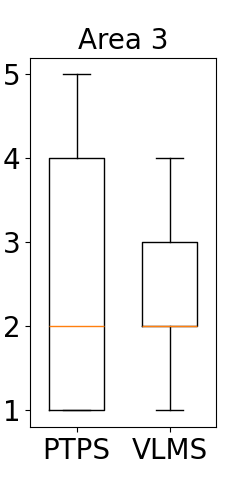}
        \caption{}
    \end{subfigure}    
    \begin{subfigure}{.3\columnwidth}
        \centering
        \includegraphics[width=.75\linewidth]{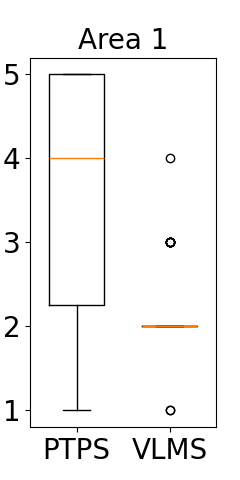}
        \caption{}
    \end{subfigure}
    \begin{subfigure}{.3\columnwidth}
    \centering
        \includegraphics[width=.75\linewidth]{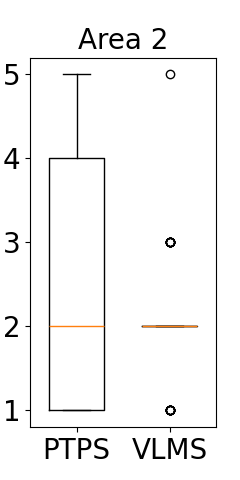}
        \caption{}
    \end{subfigure}
    \begin{subfigure}{.3\columnwidth}
    \centering
        \includegraphics[width=.75\linewidth]{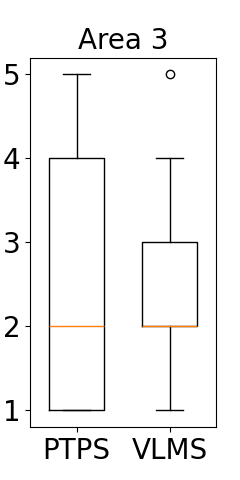}
        \caption{}
    \end{subfigure}    
    \caption{Boxplots of the overlay (a--c) and interaction (d--f) suitability ratings of participants (\ptps) and VLM (\llms{}) for the subway scenario (\autoref{fig:scenario_examples}, middle). For both questions, it can be seen that the standard deviation of \ptp{} responses is consistently larger than that of \llm{} responses. The boxplots further show that medians of both conditions frequently overlap (b, c, e, f). For areas where they do not, \ptps{} often exhibit a high standard deviation in their ratings (a).}
    \Description{The image displays a set of six box plots, labeled from (a) to (f), each comparing two groups labeled ‘PTPS’ and ‘VLMS’. All box plots show a range of data points with a central rectangle representing the interquartile range, a line inside the rectangle indicating the median, and whiskers extending to the highest and lowest values excluding outliers, which are marked as individual points. The plots are arranged in two rows of three and vary slightly in the distribution of their data points.  The boxplots further show that medians of both conditions frequently overlap (b, c, e, f). For areas where they do not, ptps often exhibit a high standard deviation in their ratings (a).}
    \label{fig:suitability_ratings}
\end{figure}

Analyzing the distributions of \llms{} and \ptps{} across areas and scenarios reveals that the standard deviation of \ptp{} responses is consistently larger than that of \llm{} responses (\qo: \ptps{} $SD=1.72$, \llms{} $SD=1.18$; \qi: \ptps{} $SD=1.74$, \llms{} $SD=1.11$).
This can also be seen in the area ratings of the subway scenario (\autoref{fig:scenario_examples}, middle). 
Its boxbplots exemplify that medians of both conditions frequently overlap (\autoref{fig:suitability_ratings}, \ref{fig:suitability_ratings}c, \ref{fig:suitability_ratings}e, \ref{fig:suitability_ratings}f). For areas where they do not, \ptps{} often exhibit an even higher standard deviation in their ratings (\autoref{fig:suitability_ratings}).

In addition, we explored whether \llms{} and \ptps{} provided the same reason when an area was deemed unsuitable, i.e., when the median rating for both groups fell below 3 - 'Neutral' (0.28 of areas of both \qo~and \qi).
Specifically, we compared the fraction when the mode of responses for questions \qroi{} was consistent across both groups.
The mode for \qro{} was identical between \ptps{} and \llms{} in 50\% of ratings (chance would be 20\%). 
For \qri, this similarity was observed in 25\% of ratings.

\subsection{Discussion}

Our findings suggest that \projectname's reasoning module is capable of assessing situations in shared social spaces not different than experienced MR users. 
When evaluating both the suitability of overlays and the appropriateness of interactions, instances of \llms{} did not assign more extreme ratings to situations than \ptps{}. 

Our analysis also revealed that \llms{} consistently assigned high ratings for overlay suitability to areas featuring any type of display ($MD=5, SD=0.72$), regardless of the context or the display's status (on or off).
In comparison, \ptps{}' assessments of display overlay suitability varied ($MD=4, SD=1.62$), showing that participants took contextual factors, such as whether the display was active, into account. 
To mitigate the influence of this bias from \llms{} on our findings, we removed all areas with displays from our analysis (affecting three areas in total).
Furthermore, we implemented a refinement in the context prompt provided to the VLM (added the sentence: ``When a monitor displays content, overlaying a virtual element on top of it is unsuitable.'').
We used the new context prompt to generate results for the MR layout study and the applications.

While our statistical analysis showed that \llms{} did not assign more extreme ratings than \ptps{}, their reasoning about the unsuitability of certain areas for UI element placement differs (with 50\%- [\qro{}] and 25\% overlap [\qri], respectively, by a 20\% coincidence rate).
It is important to mention that we did not specifically fine-tune \llms{} for reasoning responses, as our system is mainly concerned about suitability ratings.
Consequently, we anticipate that the reasoning alignment between \llms{} and \ptps{} would also increase through a fine-tuning process.


\begin{figure}[t]
    \centering
    \includegraphics[width=\linewidth]{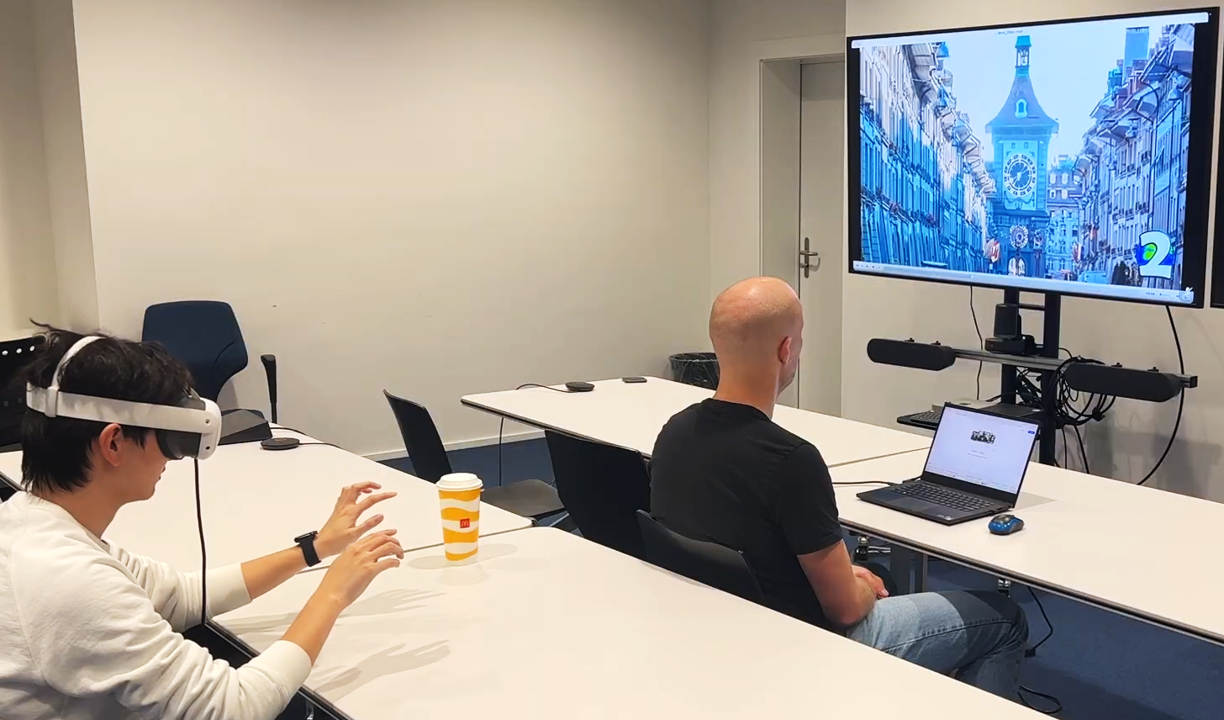} 
    \caption{Our study setup replicated a university seminar room, where the participant was sitting in the last row and another attendee was seated in the row ahead.}
    \Description{Two individuals are seated at a table in a conference room. The person on the left is wearing a virtual reality headset and holding what appears to be a controller in their hands. The person on the right is facing away from the camera, looking towards a large monitor displaying an image of an urban street with classical architecture. There is also a laptop and a disposable coffee cup on the table.}
    \label{fig:study_setting}
\end{figure}

\section{Evaluation of MR Layout Adaptation}

To evaluate if our approach generates MR layouts that better adapt to situations in shared spaces, we compared 
it with two baseline adaptation methods.
Our study thus investigated the impact of our approach on the positioning of UI elements within shared spaces, taking into account their (1)~overlay in the physical environment and (2)~assessing the ease of direct interaction with them.

\subsection{Study design}
\label{sec:exp_desing_study_2}
We used a within-subject design with two variables: \ivtask{} (2 levels: \emph{listening comprehension}, \emph{discussion}), and \ivmethod{} (3 levels: \cauit{}, \cadapt{}, \cours{}).
For each displayed UI element, we collected participants rating for its \emph{overlay-} and \emph{interaction suitability}.
Thus, we slightly adjusted the questions of the survey (overlay suitability: "Please rate the suitability of displaying the [UI element] where it was in this room."; interaction suitability: "Please rate how acceptable you found the direct interaction with the [UI element] given your surroundings and the people and objects in it.").
Responses were recorded using a 5-point Likert scale, with options ranging from 1 - "Unsuitable" to 5 - "Perfectly Suitable".
\change{Participants were asked to score the suitability considering the \textbf{FASH} factors of the user interface.
Therefore, they were introduced to these factors at the beginning of the study.}
The \ivtask{} order was fixed while \ivmethod{} orders were fully counterbalanced.

\subsubsection*{Environment}
Mimicking a shared social space, we ran the study in a seminar room of a university (\autoref{fig:study_setting}). 
The participant was seated in the last row. 
In the row before them the experimenter acted as another person attending the lecture. 
Depending on the task, they either watched the lecture or engaged with the participant. 
The lecture was played on a large screen before both of them.
\change{This environment included several \textbf{FASH}-relevant features, including functionality (display on/off), social acceptability (person facing towards/away), and health \& safety considerations (a drink that could be spilled if occluded).
These factors were the most frequently cited in our first study and are common in other real-world scenarios.}

\subsubsection*{Tasks}
\begin{figure}[t]
    \centering
    \includegraphics[width=\linewidth]{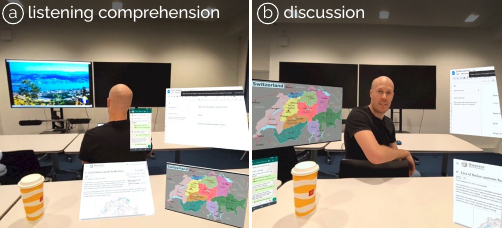}
    \caption{Participants' perspective of the adapted MR UI during the (a)~\emph{listening comprehension} and (b)~\emph{discussion} task.} 
    \Description{Two side-by-side photos of a person in a classroom setting with two computer monitors. In the first image, labeled (a) listening comp., the person is facing a monitor displaying a colorful map and another monitor with text, while in the second image, labeled (b) discussion, the person has turned to face another seat as if engaging in conversation. Both images have the individual’s face obscured for privacy. A paper with text and graphs, as well as a disposable coffee cup with stripes, are on the desk in both images.}
    \label{fig:study_task}
\end{figure}

The study involved two tasks typical of a lecture setting and purposely designed to alter the context within the seminar room.
Participants first performed a \emph{listening comprehension} task, which involved watching a geography lecture on a state of Switzerland and answering simple questions (e.g., number of inhabitants of a state) by typing the answers into a notepad widget of the MR UI (\autoref{fig:study_task}a). 
In this task, participant and the experimenter were focusing on the video lecture playing on the large screen.

Subsequently, participants performed the \emph{discussion} task.
Thus, the experimenter turned around and engaged in a conversation with the participant. They asked the participant two questions about the geography of the state (i.e., highest point of elevation, number of lakes). The participant could answer the question by scrolling through a Wikipedia page or looking at the map of Switzerland. Both were provided as widgets in the MR UI (\autoref{fig:study_task}b).

\subsubsection*{Methods}
We compared \cauit{}, \cadapt{}, and \cours{}. 
All conditions were implemented using AUIT \cite{evangelista2022auit}.
To ensure a fair comparison, all conditions were made aware of the objects (TV screens, paper cup, desk), the available free spaces, and the person present in the participant's surroundings.
Thus, we manually aligned the virtual and physical environments and represented objects as 3D bounding boxes within the AUIT optimization space.

\cauit{} places elements in a sphere around the user using the distance, field of view, look at, occlusion terms, constant view size of AUIT. In addition, we set a physical anchor for the keyboard to align with the desk.
\change{The weights assigned to the various factors are as follows: occlusion is weighted at 0.3, look at at 0.1, distance at 0.15, field of view at 0.3, and constant view size at 0.15.}
It is comparable to how virtual environments are displayed on commercial platforms such as Meta Quest or Apple Vision Pro.

\cadapt{} uses the same terms as \cauit{}, and further incorporates the interaction term described in \autoref{eq:interaction}.
To prioritize placement of elements on surfaces, the interaction suitability score $i_b$ was empirically set. 
The interaction frequency $f_v$ of each virtual element was designed to fit its functionality.
\change{The weights assigned to the various factors are as follows: occlusion is weighted at 0.2, look at at 0.1, distance at 0.1, field of view at 0.2, constant view size at 0.1, and interaction suitability at 0.3.}
The condition serves as a representative baseline for methods aligning MR UI layouts with physical surfaces, as it has demonstrated usability benefits \cite{cheng2023interactionadapt}.

\cours{} represents our system's output. To ensure a stable environment across conditions, we also use the predefined physical environment with it. 
To attain ratings from our reasoning module, we captured a photo from the position of participants with a camera and manually annotate the 2D bounding boxes for each object of the defined physical environment. 
To simulate a realistic setting, we ran a separate VLM query for each participant and used the attained ratings in the MR UI optimization.
\delete{The VLM was fine-tuned using the training data split of the online survey}.
\change{The training data split of the online survey was again added as context to the VLM.}
We used the same values for interaction frequency $f_v$ than in \cadapt{}.
\change{The weights assigned to the various factors are as follows: occlusion is weighted at 0.2, look at at 0.05, distance at 0.1, field of view at 0.2, constant view size at 0.1, overlaying suitability at 0.15 and interaction suitability at 0.2.}

\subsection{Procedure}
Participants started the study by completing a consent form and a demographic questionnaire. 
They then performed a training trial in which they familiarized themselves with the available UI elements and practiced interaction.
\change{During training, participants were introduced to the \textbf{FASH} factors and how they influence MR UI layouts in shared spaces.}
Afterwards, participants completed the conditions of the study, performing lecture and discussion tasks for each of the three adaptation methods (completing six trials in total).
Participants completed a questionnaire after each trial. 
Finally, participants ranked the three adaptation methods according to preference. 
They completed sessions in under 30 minutes.

\subsection{Participants}
\label{sec:study_participants}
We recruited 12 participants (4 female, 8 male), ages 22--29 (M=26, SD=2.1) from a local university.
They reported their frequency of using a VR/AR headset and using direct touch for MR interaction. 
For both questions, two participants mentioned using it several times a week, while eight indicated usage several times a month, and the remaining two participants reported less frequent usage.

\subsection{Results}
We analyzed the effect of \ivmethod{} across the different \ivtask{}s on \emph{overlay suitability}, \emph{interaction suitability}, and \emph{method preference}.
Due to the ordinal nature of our dependent variables, we assessed differences with a two-factor Aligned Rank Transform (ART) ANOVA.
Post-hoc comparisons were then performed using the ART-C algorithm and Bonferroni correction.

\begin{figure}
    \centering
    \begin{subfigure}{.49\columnwidth}
        \centering
        \includegraphics[width=1.0\linewidth]{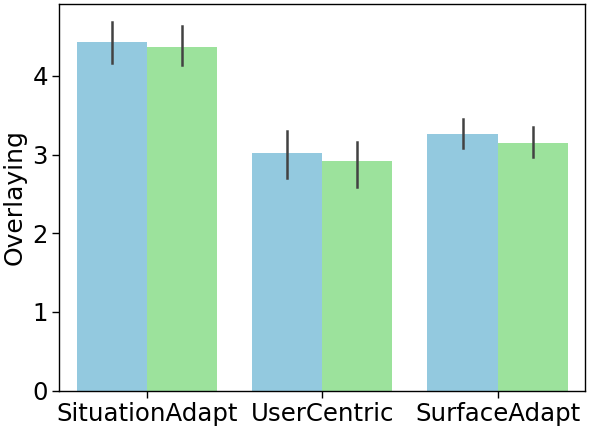}
    \end{subfigure}
    \begin{subfigure}{.49\columnwidth}
    \centering
        \includegraphics[width=1.0\linewidth]{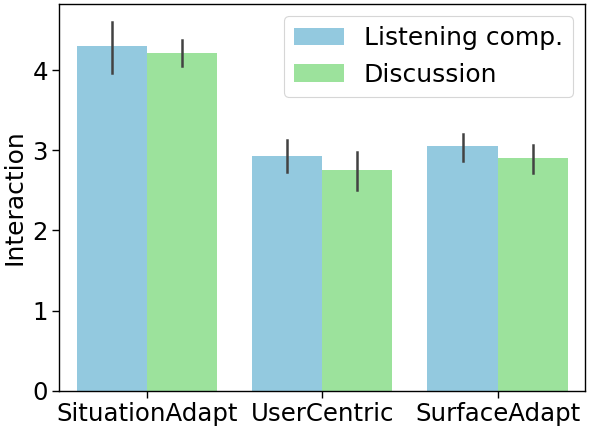}
    \end{subfigure}
    \caption{Mean and 95\% confidence interval of participant ratings per condition for overlaying- (left) and interaction suitability (right) over all UI elements and tasks.}
    \Description{The image depicts two bar graphs side by side. The left graph is labeled “Over-planning,” and the right graph is labeled “Interaction.” Both graphs have four bars each, representing different categories: “SituationAdapt,” “UserCentric,” and “SurfaceAdapt.” The bars are colored in shades of blue and green. The left graph shows a slight decrease from the first to the last category, while the right graph shows a more noticeable decrease in the same direction. Each bar has an error line indicating variability. The y-axis for both graphs ranges from 0 to 4, and the x-axis lists the categories. The top of each graph features a legend indicating that blue bars represent ‘Listening comp.’ and green bars represent ‘Discussion’.}
    \vspace{-5mm}
    \label{fig:study_results}
\end{figure}

We found a significant effect of \ivmethod{} on \emph{overlay suitability} across \ivtask{}s \anova{2}{66}{67.35}{<}{.0001}~.
Post-hoc tests have shown that \cours{} caused participants to perceive UI elements to be placed at more suitable locations in a shared space compared to \cauit{} and \cadapt{} (\pvall{<}{.0001} for both, \autoref{fig:study_results} left).
Other differences were non-significant.

Similarly, a main effect of \ivmethod{} on \emph{interaction suitability} across \ivtask{}s was found \anova{2}{66}{68.93}{<}{.0001}~.
Post-hoc analyses revealed that participants perceived UI elements as being positioned in more interaction-friendly locations within shared social spaces when using \cours{} compared to either \cauit{} or \cadapt{} (\pvall{<}{.0001} for both, \autoref{fig:study_results} right). No other significant differences were observed.

We found a main effect of \ivmethod{} on participants' preference rankings \anova{2}{66}{143}{<}{.0001}~.
Participants ranked \cours{} ($M=1.0, SD=0.0$) significantly higher than both \cadapt{} ($M=2.25, SD=0.44$) and \cauit{} ($M=2.75, SD=0.44$; \pvall{<}{.0001} for both).
We also found a statistically significant difference in ranking between \cadapt{} and \cauit{} \pval{<}{.0001}.
No other significant differences were identified.

\subsection{Discussion}

Participants reported perceiving layouts produced by \cours\ to place UI elements at locations that are more suitable in terms of overlaying a shared space.
They also perceived UI elements as being positioned in interaction-friendly locations that are suitable given the context of a shared space.
\change{Participants explained their ratings, noting that SurfaceAdapt aligns widgets with desks, making them harder to see compared to mid-air placements, and UserCentric often ignores the real-world context, frequently arranging widgets in ways that obstruct the TV or a classmate's face. 
In contrast, SituationAdapt avoids occluding important real-world areas, places interactive widgets on tables, and positions informational widgets in mid-air.}
These results suggest that \cours\ can generate MR layouts that consider the situation of the shared space surrounding the user.
Moreover, the results indicate a preference for layouts generated by our method over \cauit\ and \cadapt, highlighting the positive impact of adapting to the user's shared surroundings on layout preference. 

\section{Scenarios}

We demonstrate \projectname's ability in comprehending the context within a shared space and accordingly optimizing the placement of virtual elements across two distinct scenarios.

\begin{figure*}
    \centering
    \begin{subfigure}{.16\linewidth}
        \centering
        \includegraphics[width=1.\linewidth]{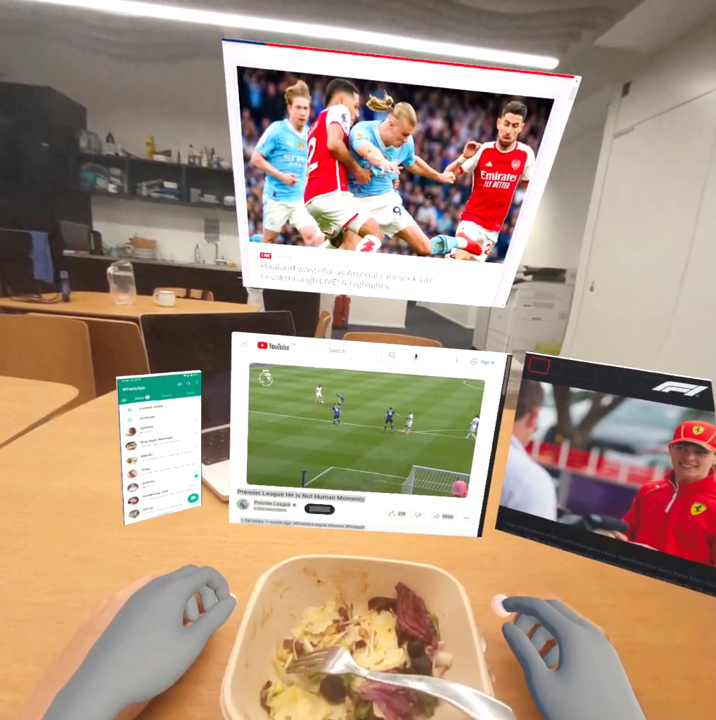}
        \caption{}
        \label{fig:app_kitchen_before}
    \end{subfigure}
    \centering
    \begin{subfigure}{.16\linewidth}
        \centering
        \includegraphics[width=1.\linewidth]{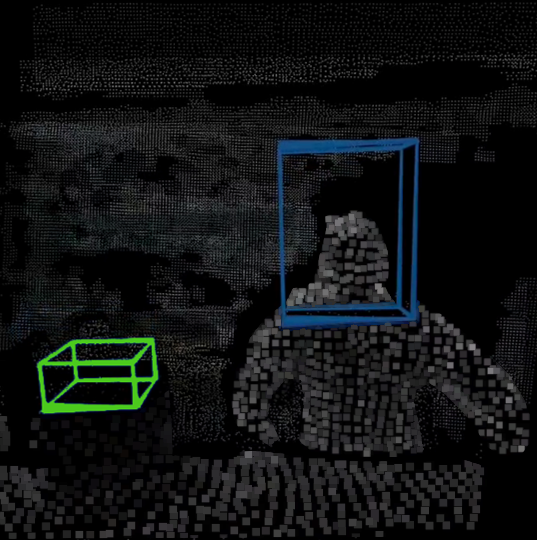}
        \caption{}
        \label{fig:app_kitchen_pointcloud}
    \end{subfigure}
    \centering
    \begin{subfigure}{.16\linewidth}
        \centering
        \includegraphics[width=1.\linewidth]{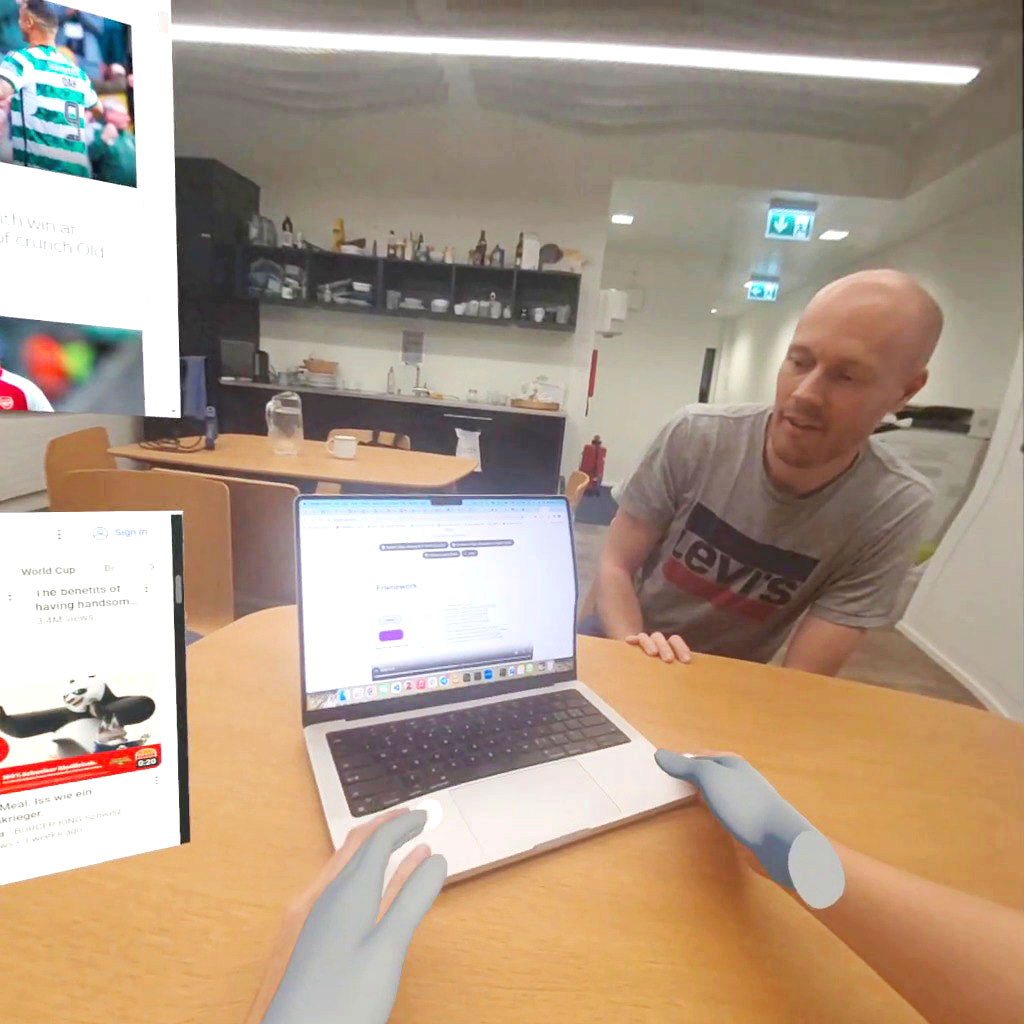}
        \caption{}
        \label{fig:app_kitchen_after}
    \end{subfigure}
    \centering
    \begin{subfigure}{.16\linewidth}
        \centering
        \includegraphics[width=1.\linewidth]{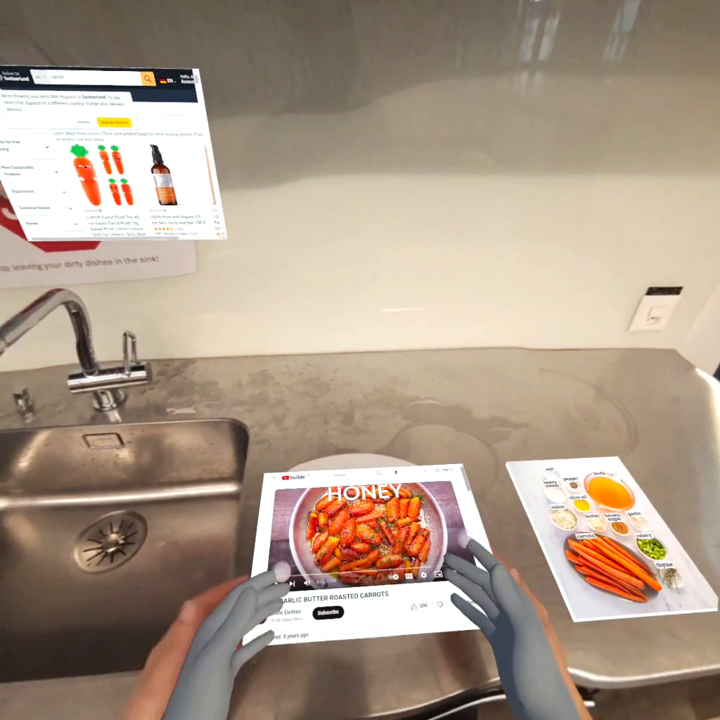}
        \caption{}
        \label{fig:app_cooking_before}
    \end{subfigure}
    \centering
    \begin{subfigure}{.16\linewidth}
        \centering
        \includegraphics[width=1.\linewidth]{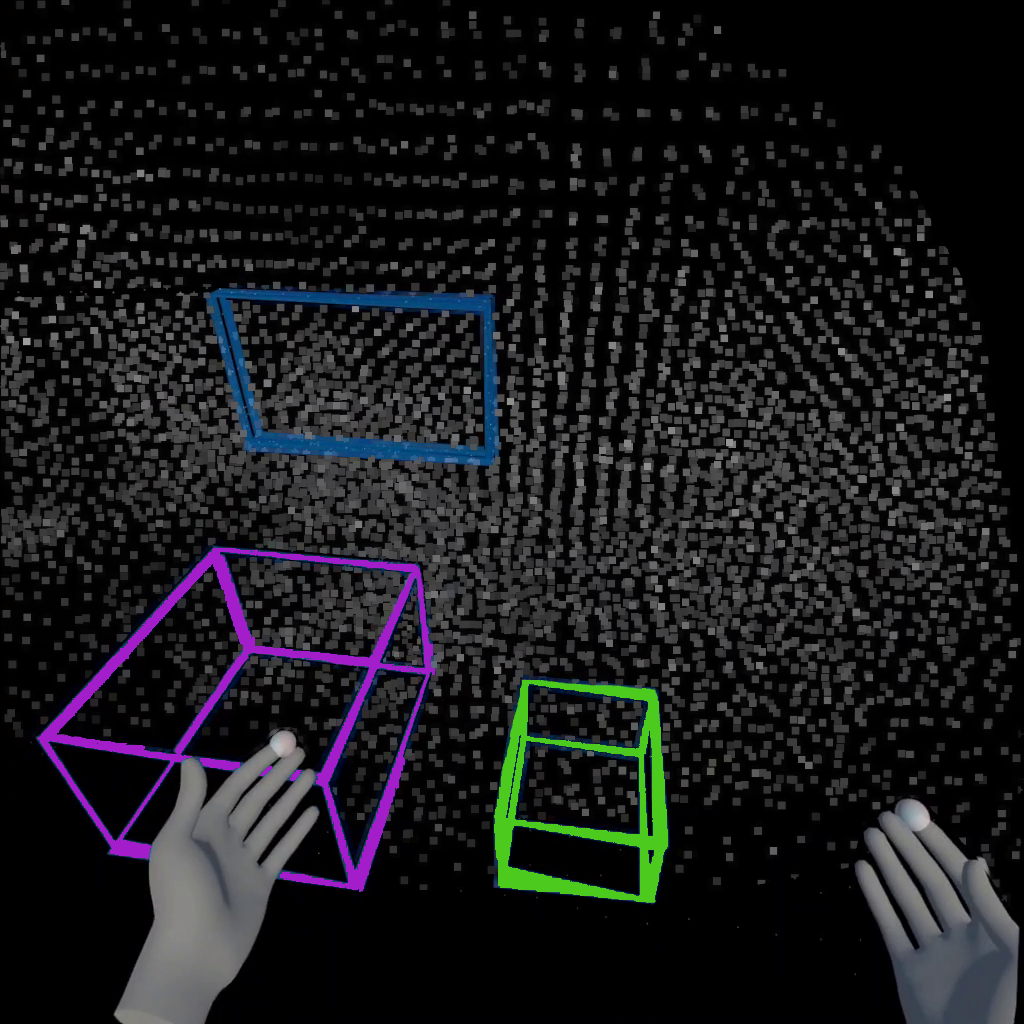}
        \caption{}
        \label{fig:app_cooking_pointcloud}
    \end{subfigure}
    \centering
    \begin{subfigure}{.16\linewidth}
        \centering
        \includegraphics[width=1.\linewidth]{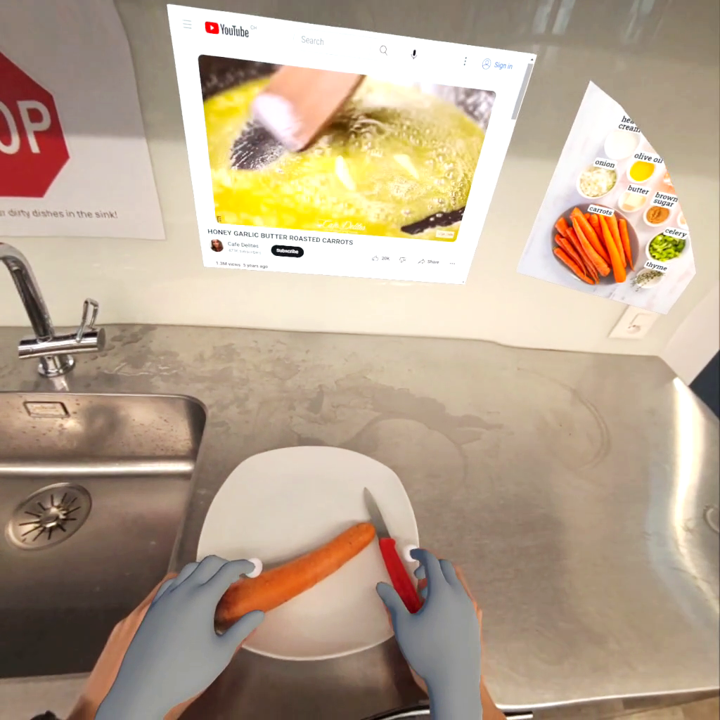}
        \caption{}
        \label{fig:app_cooking_after}
    \end{subfigure}
    \caption{We demonstrate \projname's versatility in six use-cases:
    (a)~the user browses sports news while having food,
    (b)~debug output of the perception module showing the point cloud and the detected bounding boxes for the colleague (blue) and the laptop (green),
    (c)~the virtual interface is adapted to keep the colleague and laptop unobstructed,
    (d)~the user puts the headset on finding virtual elements to occlude the plate and the warning sign,
    (e)~debug output of the perception module illustrating the point cloud and the bounding boxes for the sink (purple), plate (green), and warning sign (glue),
    (f)~ the virtual elements are adapted to keep the plate, sink and warning sign unobstructed.}
    \Description{(a): A football game displayed on a screen with an overlaid graphic of a green rectangle and a red trapezoid on the field. (b): Mostly obscured by digital noise, but there’s an overlay of a blue rectangle and a green parallelogram. (c): A person sitting at a desk with multiple computer screens; one screen has an overlaid purple trapezoid. (d): A meal tray with various food items and an overlaid orange rectangle on the tray. (e): Also obscured by digital noise, with overlays of pink and green geometric shapes. (f): Another view of the meal tray from (d), this time with an overlaid yellow parallelogram on the tray.}
    \vspace{-3mm}
\end{figure*}

\subsection{Discussion over lunch}


We demonstrate \projectname within a cafeteria setting.
In this context, the user takes a break from work and is having lunch. 
While eating, the user watches sports videos through a virtual browser, surrounded by various virtual widgets such as sports news and messaging apps (\autoref{fig:app_kitchen_before}).
According to this initial context, all virtual elements are placed around the user, optimizing their visibility and spatial distribution.
After a while, the user's colleague comes to the table, activates the laptop and starts a chat with the user.
Our perception module detects the colleague and the laptop and fits the respective 3D bounding boxes (\autoref{fig:app_kitchen_pointcloud}).
The reasoning module detects that the colleague faces the user and that the laptop is turned on and provides suitability ratings.
Based on these ratings, our optimization module dynamically adjusts the layout of virtual elements.
All virtual elements are re-positioned away from the colleague and prevented from overlaying the laptop, ensuring the content under discussion remains unobstructed (\autoref{fig:app_kitchen_after}).
This scenario demonstrates how \projectname adapts an MR UI according to the factors of 'Functionality' and 'Social acceptability'.

\subsection{Preparing a meal}

We demonstrate \projectname in the context of a food preparation scenario.
While this scenario does not feature other people, we chose it to demonstrate the usefulness of our system's adapted layouts in single-user workspaces.
Within this context, the user first browses groceries, online recipes, and cooking videos in their office, where all virtual elements are placed on surfaces optimized for touch interaction. 
Once the user arrives in the kitchen and puts the headset on, the widgets adhere to physical surfaces according to their initial objective of facilitating interaction.
As a result, the virtual elements occlude important physical objects in the kitchen, including plate and knife on the counter as well as a warning sign on the wall (\autoref{fig:app_cooking_before}). 
Our perceptual module identifies the respective objects in the environment, and extracts their 3D bounding boxes (\autoref{fig:app_cooking_pointcloud}). 
Subsequently, the reasoning module detects the best locations for placing virtual elements in the environment. 
Based on the ratings, our optimization module dynamically changes the layout to keep the virtual elements visible and prevent occlusion of warning sign, knife and plate (\autoref{fig:app_cooking_after}). 
This scenario exemplifies how \projectname considers the factors of 'Health\&Safety' and 'Functionality' when adapting MR user interfaces.

\section{Discussion \& Future Work}
We developed \projectname\ to enable immersive interfaces to adapt to the situational context in shared spaces. 
In the following, we discuss limitations of our work as well as remaining open questions related to the research direction in general.

\subsubsection*{Perception of surroundings}
While the implementation of our perception module is a means to an end, we still want to discuss its limitations.
With RTAB-Map~\cite{noauthor_rtab-map_2023}, we build on top of a traditional SLAM approach that was designed to map and navigate static environments without considering moving objects.
To overcome this limitation, we manually re-initialized it with each contextual change, allowing us to retain a new 3D map per situation.
Future research should use Dynamic- \cite{wang2007simultaneous} or Semantic SLAM approaches \cite{bowman2017probabilistic} to track moving objects and people in the environment.

Our implementation of the perception model is also limited by the set of objects that YOLOv3 can recognize, which furthermore do not include large surfaces like walls.
Future work should rely on objects detection methods that span a vast set of categories \cite{zhou2022detecting} and fuse these information with real-time segmentation approaches~\cite{wang2022rtformer} to also gain an understanding of the surfaces in the scene.

In our current implementation, the user themselves communicates a contextual change via button press to the perception module.
Future research should investigate how such a change could automatically and reliably be detected.
One possible strategy could be to leverage positions and confidence values of a Semantic SLAM to discern when an object becomes relevant to the user's context.

Furthermore, SDKs of future MR headsets should grant developers access to environment reconstruction and understanding features, facilitating the creation of context-aware applications without needing external hardware or redeveloping localization, mapping, and semantic understanding functionalities.

\subsubsection*{User study}

We evaluated \projectname\ in a single scenario, in which the context in a simulated lecture changed from watching a video to discussion with a classmate.
However, the context of real-world shared spaces is typically more dynamic, including multiple individuals who may be strangers or friends.
In addition, our user study only manipulated the shared space considering 'Health \& Safety', 'Function' and 'Social acceptability' of the \textbf{FASH} factors.
Future research should explore the functionality of our system in real-world shared spaces and also investigate how users perceive UI adaptations caused by all of the \textbf{FASH} factors.


\subsubsection*{VLMs for UI adaptation}

In our reasoning module, VLMs utilize users' field of view as input alongside pre-designed prompts to attain human-like suitability ratings.
However, users may have additional information beyond the current field of view when assessing the suitability of placing virtual elements in a shared space. 
For instance, users might prioritize overlaying virtual elements over strangers in public spaces while preferring to keep friends unobstructed. 
Inferring these relationships solely from images is unfeasible.
We believe that prompting VLMs with user-specific historical data and information could help construct a context for each user and thus facilitate personalized adaptive user interfaces.


As current MR devices are designed mostly for indoor use, all scenarios in our survey were focusing on indoor environments. 
Initial tests in outdoor settings revealed that the VLM frequently took into account factors human evaluators considered as insignificant. 
Future research should investigate how VLMs comprehend shared outdoor environments and explore methods to improve their ability to accurately assess these settings.
With AR glasses soon to become a mainstream consumer device, this would enable MR layout adaptation to shared outdoor spaces.

In the broader context of general HCI, we believe that our research sheds light on whether AI models are able to simulate user behavior, contributing to the discourse on AI versus human reasoning.
In our study, we found an interesting dichotomy in that the VLM was capable to provide ratings not different than experienced MR users, however, it struggled to provide reasoning that aligned with the rationale of these users.
This aligns with findings from other studies indicating that LLMs can produce artificial responses to open-ended survey questions \cite{schmidt2024simulating}.
Future research should dive deeper into validating if AI models can simulate human participants in the context of user evaluations \change{and further investigate the differences between human and AI reasoning}.

\section{Conclusion}

We have presented \projectname, an end-to-end system that considers social and environmental factors in optimizing UIs for Mixed Reality in shared spaces. 
\projectname perceives the physical environment with real-time object detection and 3D mapping, then reasons about the suitability of placing virtual elements with a VLM, and optimizes the MR interface accordingly.

To validate our approach, we conducted an online survey where we compared VLM responses to those of experienced MR users in terms of understanding the context of shared spaces.
Results suggest that the VLM judged the situations not different than participants.
We then evaluated the suitability of the MR layouts generated by \projectname  during a lecture scenario and compared it with two baseline approaches.
We found that participants rated \projectname's layouts as more suitable and fitting within the situated context of the shared space.

We believe that our approach contributes an important step towards truly context- and situation-aware MR systems, enabling their adaptation to the nuances of shared social settings.
We argue that this will be key to enabling MR device use in mobile settings beyond controlled home and office spaces.

\begin{acks}
\change{
We would like to express our sincere gratitude to Max Möbus for his assistance and support with the statistical analysis of our studies.
}
\end{acks}

\balance
\bibliographystyle{ACM-Reference-Format}
\bibliography{citations}

\appendix
\section{Context Prompt of LLM}
\label{app:context-prompt}
We utilized the following prompt to establish the context for the Large Language Model (LLM). This example focuses on setting the context for overlay suitability, whereas the prompt for interaction suitability was similar.

\begin{enumerate}[leftmargin=0.3cm]
    \item []{"You will mimic a participant of a survey in which participants had to rate the suitability of Mixed Reality layouts that overlay User Interfaces onto parts of the real world. Thus, you will rate the suitability of directly interacting with virtual UI elements that you imagine be placed on each highlighted area of an image.All virtual elements would only be visible to you, not to other people in the image. All virtual elements would not obstruct the view of other people or light. The people you can see in the image are someone else, not yourself. You will rate the suitability of each area on a score that ranges from 1 to 5 where 1 means 'unsuitable', 2 means 'somewhat unsuitable', 3 means 'neutral',  4 means 'somewhat suitable' and 5 means 'suitable'. 
  
    You will be asked to give the primary reason for your choice of suitability. Optional reasons are: functionality, social, health \& safety, aesthetics, and other. Functionality means: the UI element hinders the functionality of the physical object. Social acceptability means: looking at or interacting with the UI element would be socially inappropriate. Health \& Safety means: the UI element occludes safety critical information or may lead to sanitation issues during interaction. Aesthetics means: the UI element impairs the visual appeal of the physical surroundings. Other means: your primary reason is not covered in the list above. 
    
    To improve your ability to imitate a participant, you will be shown images they have evaluated and receive information about the median and standard deviation of their ratings for the highlighted areas of these images. Please take these ratings into account when judging new images."
}
\end{enumerate}

\noindent
The following prompt was utilized to provide the LLM with an understanding of how a group of users evaluated specific areas of a certain image (the numerical data is illustrative).

\begin{enumerate}[leftmargin=0.3cm]
    \item []{'Participants of a survey provided the following median responses along with standard deviations for the direct interaction suitability of the areas in this image: area 1: median 2.0, standard deviation 1.74; area 2: median 1.0, standard deviation 1.52; area 3: median 4.0, standard deviation 1.78;'}
\end{enumerate}

\end{document}